\documentclass[pra,showpacs,twocolumn,superscriptaddress]{revtex4-1}
\usepackage{amsfonts}
\usepackage{amsmath}
\usepackage{amssymb}
\usepackage{mathtools}
\usepackage{graphicx}
\usepackage{epsfig}
\usepackage{bm}
\usepackage{enumerate}
\usepackage{multirow}
\usepackage{amsthm}
\usepackage{tabularx}
\usepackage{hyperref}
\usepackage{subcaption}

\newcommand{\ket}[1]{\left| #1 \right\rangle}

\newcommand{\bracket}[2]{\left\langle #1 | #2 \right\rangle}
\newcommand{\proj}[1]{| #1 \rangle \langle #1 |}

\renewcommand{\arraystretch}{1.2}

\makeatletter
\renewcommand*\env@matrix[1][\arraystretch]{%
  \edef\arraystretch{#1}%
  \hskip -\arraycolsep
  \let\@ifnextchar\new@ifnextchar
  \array{*\c@MaxMatrixCols c}}
\makeatother

\newcolumntype{Y}{>{\centering}X}

\begin{document}


\title{Recursive multiport schemes for implementing quantum algorithms
with photonic integrated circuits}




\author{Gelo Noel M. Tabia}
\email[Electronic address: ]{gelo.tabia@ut.ee}
\affiliation{Institute of Computer Science,
University of Tartu,
J. Livii 2, 50409, Tartu, Estonia}

\begin{abstract}

We present recursive multiport schemes for implementing quantum
Fourier transforms and the inversion step in Grover's algorithm
on an integrated linear optics device. In particular, each scheme
shows how to execute a quantum operation on $2d$ modes using
a pair of circuits for the same operation on $d$ modes.
The circuits operate on path-encoded qudits and
realize $d$-dimensional unitary transformations on these states
using linear optical networks with $O\left(d^2\right)$ optical elements.
To evaluate the schemes against realistic errors, we ran
simulations of proof-of-principle experiments using a simple
fabrication model of silicon-based photonic integrated devices that
employ directional couplers and thermo-optic modulators for
beam splitters and phase shifters, respectively. We find that
high-fidelity performance is achievable with our multiport
circuits for $2$-qubit and $3$-qubit quantum Fourier transforms, and
for quantum search on four-item and eight-item databases.

\end{abstract}


\maketitle

\section{Introduction}
\label{sec:introduction}

Linear optics with single photon sources and detectors provides a
promising candidate for establishing efficient and scalable universal
quantum computation \cite{klm2001, kok2007}.
Some specific advantages of optical implementations are the
robustness of photons against decoherence and ultrafast optical
processing. Meanwhile, current limitations include low efficiencies
in photon creation and measurement, and the difficulty of storing
photonic quantum information in a quantum memory.
However, the main technical challenge with the scalability of linear
optical devices is the considerable overhead necessary for realizing
multi-qubit gates. Moreover, maintaining phase stability between
optical modes often requires active checking and calibration---a task
that clearly becomes more demanding the larger the circuit gets.
Both of these issues are addressed by integrated photonic technology.

A photonic integrated circuit (PIC) is a multiport device consisting
of an integrated system of optical elements embedded onto a single
chip using a waveguide architecture \cite{marshall2009, thompson2011}.
Because PICs are compact and designed to have inherent phase
stability, they offer the potential for a truly scalable optical
quantum computer. They are also fully compatible with electronic
devices and fiber optic systems, which can lead to increased
functionality.

There has been much progress made in PIC-based approaches to
quantum optics: recent experiments have demonstrated linear optical
quantum gates \cite{politi2008, laing2010},
multiphoton entanglement \cite{matthews2009, shadbolt2011}
boson sampling experiments \cite{aaronson2011,crespi2013, broome2013},
quantum walks in optical arrays \cite{owens2011, sansoni2012, mataloni2013},
and simulation of quantum systems \cite{lanyon2010, aspuruguzik2012}.
More generally, PICs offer a natural platform for conducting
experiments on quantum systems with higher-dimensional
Hilbert spaces \cite{tabia2012, schaeff2015}.

In this paper, we describe recursive multiport schemes for
carrying out quantum operations on a PIC. By recursive we mean that
two copies of the $d$-dimensional circuit are used to construct
the $2d$-dimensional version. Formally, a multiport circuit
represents a decomposition of a unitary transformation into a
network of single-qubit gates acting on adjacent modes, which on a
PIC is mapped onto a sequence of beam splitters and phase shifters.
It is crucial to note that in our schemes, the photons represent
multi-rail qudits, not dual-rail qubits. This is an important
distinction since existing methods for performing entangling gates
on dual-rail qubits are not very scalable. However, our circuits
act on single qudits, therefore, such entangling gates are
not needed.

We must emphasize, though, that linear optical implementations
on single photonic qudits are inherently unscalable for quantum
computation since they require an exponential number of
optical modes. The only known way for performing scalable
quantum computation with linear optics requires multi-photon
quantum interference and and measurement-based nonlinearity,
such as in the KLM proposal \cite{klm2001}. Nevertheless, there are
several contexts in which our recursive circuits may find suitable
application, for instance, within a larger architecture of
entangled qudits \cite{joo2007} or when using QFT as a
verification tool in boson sampling \cite{tichy2014},
the latter having been experimentally demonstrated
by Carolan, \emph{et al.} \cite{carolan2015}.

Here, we consider circuits for two important families of
unitary transformations: quantum Fourier transform, which is an
important subroutine in many quantum algorithms, is covered in
Section~\ref{sec:qft}, and inversion about the mean, which is a
key step for the iterations in Grover's algorithm, is covered
in Section~\ref{sec:grovInv}. In both cases, we provide a recipe
for constructing a linear optical network that uses a total of
$O\left(d^2\right)$ beam splitters and phase shifters
for realizing a quantum operation on $d$ modes.
The basis for these constructions are efficient matrix
factorizations of the $2d$-dimensional unitary operators into a
product of a matrix consisting of two copies of a $d$-dimensional
version of the same unitary operator and some sparse matrices
with $2\times 2$ blocks on the diagonal.
Thus, our schemes provide a systematic way of
constructing larger circuits from smaller ones, which is quite
beneficial for scaling these quantum operations to higher dimensions.

To demonstrate the practical viability of our multiport scheme,
we also conducted simulations of experiments on the circuits for
quantum Fourier transform and Grover's algorithm using
a fabrication model that incorporates realistic errors
in the beam splitters and phase shifters based on wafer-scale
testing data on PIC components \cite{mower2014}. The simulation
results are discussed in Section~\ref{sec:mCircSim}.

\section{Quantum Fourier transforms}
\label{sec:qft}

Many known quantum algorithms that exhibit exponential speedup
over their classical counterparts make use of quantum Fourier
transform (QFT), which describes a discrete Fourier transform on
quantum mechanical amplitudes \cite{nc2000}. Typically, the speedups
come from performing quantum phase estimation, which involves finding
approximate eigenvalues of a unitary operator, and it involves an
inverse QFT. It is especially helpful in solving interesting problems
like prime factorization and discrete logarithms \cite{shor1997}.

There have been some recent experiments that realize QFT
with optical multiport circuits, in particular, the four-mode version
by Laing \emph{et al.} \cite{laing2012}, which was performed with
path-and-polarization encoded states in bulk optics, and
a six-mode version used in the study of inequivalent classes of
complex Hadamard matrices by Carolan, \emph{et al.} \cite{carolan2015}.
These examples provide evidence for the considerable interest in
developing practical linear optical methods for implementing QFT.

In this section, we describe a recursive multiport scheme for
realizing QFT with integrated linear optics. The key ingredient for
the matrix decomposition involved is the unitary operator given by the
direct sum of two of Fourier matrices.

In order to facilitate the description of multiport circuits,
we shall use the following special notation throughout this paper.
A multiport circuit $C$ with $d$ optical modes is
called $C(1,2,\ldots, d)$. The modes are labeled $1$ to $d$
from top to bottom. Optical elements are organized from left
to right according to the sequence they appear in the circuit.
Those that can be performed in parallel are enclosed
in square brackets.

We will denote the optical elements as follows:
\begin{enumerate}
\item
$B_\epsilon(i,j)$ refers to a beam splitter with
reflectivity $\epsilon$ acting on modes $i$ and $j$.
Our convention here is to choose the overall phases so that
in terms of the mode operators we have
\begin{equation}
\begin{pmatrix}
a^\dag_{i,\text{out}} \\
a^\dag_{j,\text{out}}
\end{pmatrix}
=
\begin{pmatrix}
\sqrt{\epsilon} & \sqrt{1-\epsilon} \\
\sqrt{1-\epsilon} & -\sqrt{\epsilon}
\end{pmatrix}
\begin{pmatrix}
a^\dag_{i,\text{in}} \\
a^\dag_{j,\text{in}}
\end{pmatrix}.
\end{equation}
For simplicity, $B(i,j)$ denotes an equal beam splitter
$(\epsilon = 1/2)$ between modes $i$ and $j$.
\item
$S(i,j)$ refers to a swap operation between modes $i$ and $j$.
With the convention above, this is the same as having
a beam splitter with $\epsilon = 0$.
\item
$P_\theta(i)$ refers to a phase shifter on mode $i$, which means
the amplitude for mode $i$ is multiplied by $e^{i\theta}$.
\end{enumerate}

\begin{figure}[t]
\centering
\includegraphics[scale = 0.3]{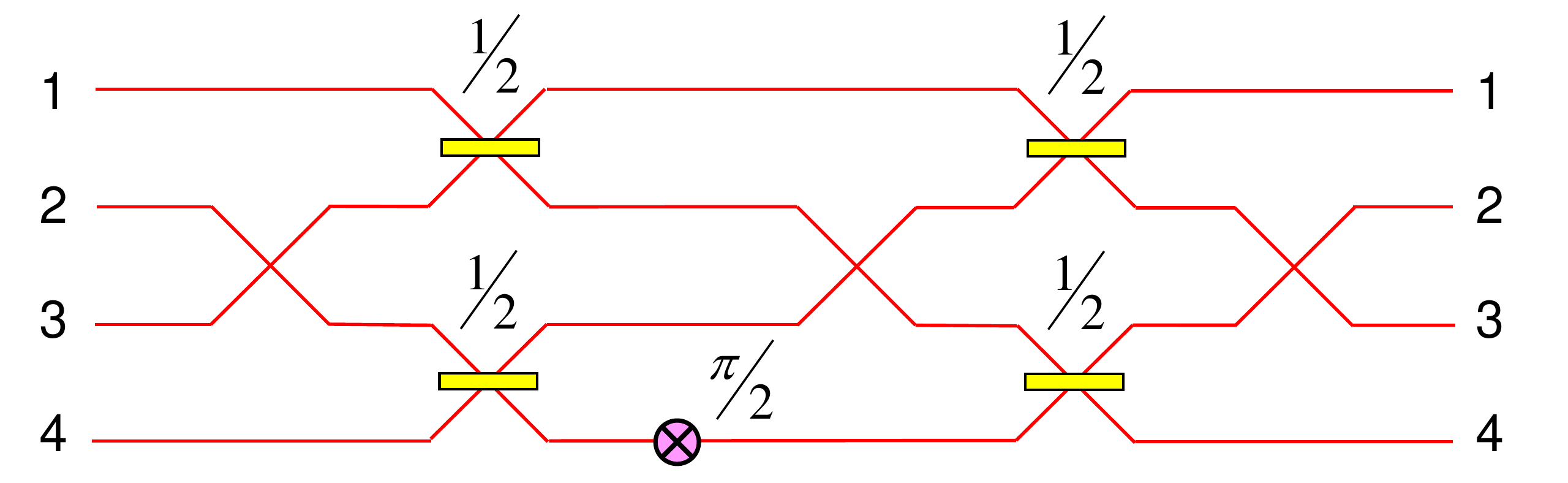}
\caption{(Color online) The optical multiport circuit for 2-qubit QFT $F_4$.}
\label{fig.quantumFourierdim4}
\end{figure}

Using our multiport circuit notation, the single-qubit QFT, or
the Hadamard gate, is just $F_2(1,2) = B(1,2)$.
For $2$-qubit QFT, the multiport circuit is given by
\begin{align}
\nonumber
F_4(1,2,3,4) &= [S(2,3)] \> [B(1,2) \> B(3,4)]
            \left[P_{\frac{\pi}{2}}(4)\right] \\
&\qquad  \> [S(2,3)] \> [B(1,2) \> B(3,4)]
						 \> [S(2,3)].
\label{eq:2-qft}
\end{align}
The circuit for $F_4$ is illustrated in
Fig.~\ref{fig.quantumFourierdim4}.

Let $d = 2^m$ be the number of modes for the $m$-qubit QFT
multiport circuit. Using $F_4$ we can construct the
$3$-qubit QFT $F_8$ as follows:
\begin{align}
\nonumber
F_{8}&(1,2,\ldots, 8) = [S(2,3) \> S(4,5) \> S(6,7)]
             \> [S(3,4) \> S(5,6)] \\
\nonumber
&\qquad [S(4,5)] \> [F_4(1,2,3,4) \> F_4(5,6,7,8)] \\
\nonumber
&\qquad \left[P_{\frac{\pi}{4}}(6) \> P_{\frac{\pi}{2}}(7)
          \> P_{\frac{3\pi}{4}}(8) \right]
          \> [S(4,5)]  \>  [S(3,4) \> S(5,6)] \\
\nonumber
&\qquad   [S(2,3) \> S(4,5) \> S(6,7)]\\
\nonumber
&\qquad	[B(1,2) \> B(3,4) \> B(5,6) \> B(7,8)] \\
\nonumber
&\qquad   [S(2,3) \> S(4,5) \> S(6,7)]
          \> [S(3,4) \> S(5,6)] \\
&\qquad   [S(4,5)]
\label{eq.3-qft}
\end{align}

\begin{figure*}[t]
\centering
\includegraphics[scale = 0.28]{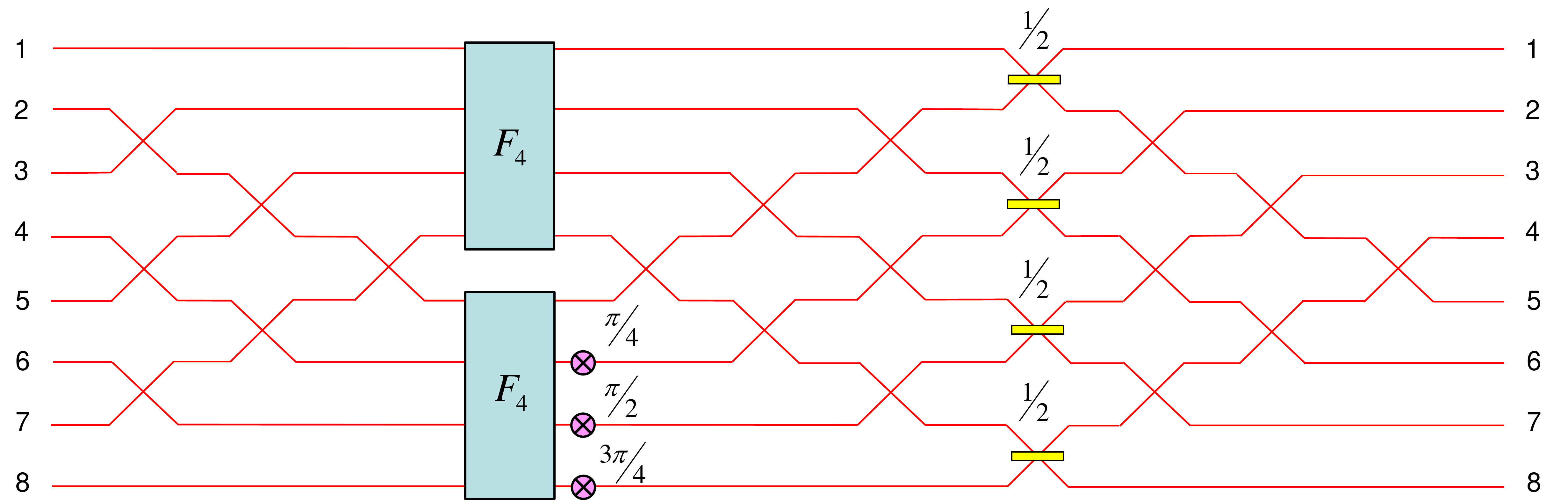}
\caption{(Color online) Implementing 3-qubit QFT using a pair of multiport
circuits for 2-qubit QFT.}
\label{fig.quantumFourierdim8}
\end{figure*}

Fig.~\ref{fig.quantumFourierdim8} shows a diagram for
how circuits for $F_4$ is used for performing the
$3$-qubit QFT $F_8$.

To describe how to build $F_{2d}$ from $F_d$, it is
convenient to define the shuffle operation $\Sigma$ that
performs the following permutation on $2d$ modes:
\begin{align}
\nonumber
(1,2, &\ldots, 2d) \mapsto  \\
&  (1,d+1,2,d+2,\ldots, k, d+k, \ldots, d, 2d ).
\end{align}
Let $\Sigma^{-1}$ denote the inverse permutation.

Observe that $\Sigma(1,2,\dots,2d)$ can be realized with
a PIC using $\frac{d(d-1)}{2}$ swap gates:
\begin{align}
\nonumber
\Sigma&(1,2,\ldots, 2d) = [S(d,d+1)] [S(d-1,d) \> S(d+1,d+2)] \\
\nonumber
&\qquad [S(d-2,d-1) \> S(d,d+1)\> S(d+2,d+3)] \cdots \\
&\qquad [S(2,3) S(4,5) \cdots S(n-2,n-1)].
\label{eq.shuffleSigma}
\end{align}
Thus,
\begin{align}
\nonumber
F_{2d}&(1,2,\ldots, 2d) = \left [\Sigma^{-1}(1,2,\ldots, 2d) \right] \\
\nonumber
&\qquad [F_d(1,2,\ldots,d) \> F_d(d+1,d+2,\ldots,2d)] \\
\nonumber
&\qquad \left[P_{\frac{\pi}{d}}(d+2) \cdots P_{\frac{k\pi}{d}}(d+k+1)
           \cdots P_{\frac{(d-1)\pi}{d}}(2d) \right] \\
\nonumber
&\qquad \left[\Sigma(1,2,\ldots, 2d) \right] \\
\nonumber
&\qquad [B(1,2) \> B(3,4)\cdots B(2d-1,2d)] \\
&\qquad \left[\Sigma^{-1}(1,2,\ldots, 2d)\right]
\end{align}

Let $N(F_d)$ denote the number of optical elements needed for
the $d$-dimensional QFT multiport circuit.
With $N(F_2)=1$ and $N(F_4) = 8$, we get the following recursive
formula for the circuit size:
\begin{equation}
N(F_{2d}) = 2 N(F_d)
+ \frac{3d\left(d - 1\right)}{2} + d
 + \left(d - 1\right).
\end{equation}
The first term in the sum on the right-hand-side
corresponds to a pair of QFT circuits on $d$ modes, the second term
counts the total number of swap gates, the third term
counts the equal beam splitters, and the last term counts the
phase shifters. Expressed as a function of $d$, we have
\begin{equation}
N(F_d) = \frac{3d^2 + d\left(\log_2{d} - 7\right)}{4} + 1
\end{equation}
so the total number of elements in the multiport
circuit is quadratic in the number of modes, as expected.

The recursive scheme works not just for qubits but for any
QFT circuit with an even number of modes. More precisely,
given an initial QFT circuit $F_n$,
the scheme can be used to implement QFT for $d = n 2^k$,
where $k$ is a positive integer.

The scheme is essentially equivalent to a PIC translation
of an important matrix factorization of Fourier matrices, initially
discovered by Gauss \cite{strang1993} and a precursor to
algorithms that perform fast Fourier transform \cite{cooley1965}:
\begin{equation}
F_{2d} = \frac{1}{\sqrt{2}}
\begin{pmatrix}
I_d & D_d \\
I_d & -D_d
\end{pmatrix}
\begin{pmatrix}
F_d & 0 \\
0 & F_d
\end{pmatrix}
P
\label{eq.fftdec}
\end{equation}
where $F_d$ is the $d$-dimensional Fourier matrix,
$I_d$ is the $d$-dimensional identity matrix,
$D_d = \mathrm{diag}\left(1,\omega,\ldots, \omega^d\right)$ is
a diagonal matrix with $\omega = e^{2\pi i/d}$, and
$P$ is the $2d \times 2d$ permutation matrix
that maps the column vector
$\vec{v} = \left( v_1, v_2, \ldots, v_{2d} \right)^T$ into
\begin{equation}
P \vec{v} = \left( v_1, v_3, \ldots, v_{2d-1}, v_2, v_4,
             \ldots, v_{2d} \right)^T,
\end{equation}
that is, it shuffles components of $\vec{v}$ such that
the first half involves components with odd indices and
the second part involves the  even ones. Note that $P$ is actually
the matrix representing the permutation $\Sigma^{-1}$.

It is important to note that linear optical implementations
of QFT have been previously explored
by T\"{o}rm\"{a}, \emph{et al.} \cite{torma1996},
and Barak and Ben-Aryeh \cite{barak2007}.
T\"{o}rm\"{a}, \emph{et al.} examine the sufficient number of
beam splitters needed for totally symmetric mode couplers,
of which the discrete Fourier transform is a special case.
Their calculations show that for a $d$-mode circuit, $(d \log_2 d)/2$
beam splitters are sufficient. On the other hand, Barak and Ben-Aryeh
describe a particular linear optical scheme for QFT based on
the Cooley-Tukey algorithm.

A crucial difference between our scheme and these other approaches
is that we are restricted to beam splitters that operate only
on adjacent modes, since this is a limitation on PICs.
This is why our scheme generally requires more beam splitters,
in order to perform those additional swap operations.
Both of the previous schemes were designed with bulk optics in mind,
so they do not consider this restriction.

For QFT, all three approaches are formally equivalent through
Eq. \ref{eq.fftdec}. In fact, the same number of equal
beam splitters is used in all schemes; it is the number of
phase shifters that vary.

T\"{o}rm\"{a}, \emph{et al.} provide a general formula for the matrix
factorization they used but it does not completely specify where
phase shifts are actually needed, since they are mostly concerned
with counting the beam splitters.

Barak and Ben-Aryeh describe a specific Cooley-Tukey factorization
of $d$-dimensional Fourier matrices into $\log_2 d$
unitary operations that contain $d/2$ pairs of beam splitters and
phase shifters, i.e., the phase shifts are always coupled to a beam
splitter. We note that a recent experiment
by Crespi, \emph{et al.} \cite{crespi2015} for testing the quantum
suppression law \cite{tichy2014} implements the Barak and Ben-Aryeh
circuits for four- and eight-mode QFT on a 3-D PIC.

In contrast, our scheme employs $d-3$ phase shifters for
$d$-mode QFT, so we achieve a modest savings on phase shifters.
For example, the $3$-qubit QFT circuit of Barak and Ben-Aryeh uses
12 phase shifters but ours uses only 5. It may be worth mentioning
that if one actually implements our QFT circuit with bulk optics,
or a 3-D PIC such as in Ref. \cite{crespi2015},  where the swap gates do
not involve beam splitters, our circuit is slightly more efficient
because it requires fewer phase shifters.

\section{The inversion step in Grover's algorithm}
\label{sec:grovInv}

Grover's algorithm \cite{grover1996} describes a quantum algorithm
for searching an unsorted database of $d$ items using
$O(\sqrt{d})$ calls to an oracle, which offers a quadratic speedup
over known classical methods. In the most basic scenario, we have a
database with $d$ items and we are supplied with a quantum oracle
that can mark the solution to the search problem by shifting the
phase of the solution's register. The goal of the algorithm is
to find the solution using the smallest number of queries.

To start, we prepare the equal superposition state
\begin{equation}
\ket{\psi} = \frac{1}{\sqrt{d}} \sum_{x =1}^{d} \ket{x},
\end{equation}
where $d = 2^m$ is the number of basis states $\ket{x}$
for $m$ qubits.

Grover's algorithm is then characterized by the repeated use of a
quantum subroutine known as the Grover operator
\begin{equation}
G = \left(2\proj{\psi} - I \right) O,
\end{equation}
where $O$ is a query to the oracle and
\begin{equation}
W = 2\proj{\psi} - I
\end{equation}
is often called the inversion about the mean. We shall refer to
$W$ as Grover inversion. In this section, we describe a recursive
multiport scheme for implementing Grover inversion on a PIC.
In particular, the multiport circuit for searching $d$ items is
utilized as a building block to the circuit for
searching $2d$ items.

If we consider a search problem with a unique solution, the
oracle can be realized by a single $\pi$-phase shift on
the optical mode corresponding to the item to be marked.

\begin{figure}[t]
\centering
\includegraphics[scale = 0.3]{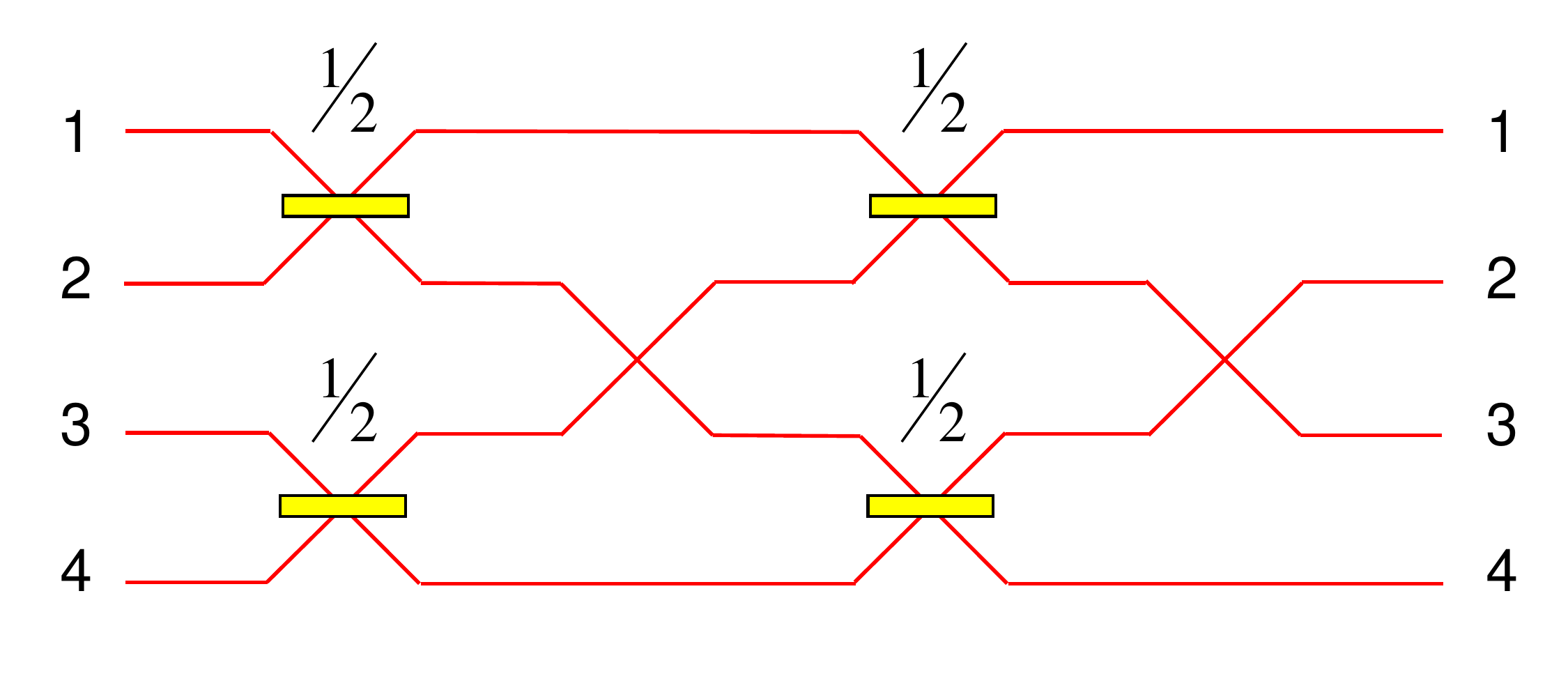}
\caption{(Color online) The optical multiport circuit for the unitary $V_4$. }
\label{fig.groverIterV4}
\end{figure}

To describe the main result, it is helpful to first consider
the unitary transformation $V_d$,
which involves a relatively simple network of equal
beam splitters on $d$ modes. Using the multiport notation
presented in Section \ref{sec:qft}, we have
\begin{align}
\nonumber
V_4(1,2,3,4) &= [B(1,2) \> B(3,4)] \> [S(2,3)] \\
& \qquad  [B(1,2) \> B(3,4)]  \> [S(2,3)] .
\end{align}
The multiport circuit for $V_4$
is depicted in Fig.~\ref{fig.groverIterV4}.

\begin{figure*}[t]
\centering
\includegraphics[scale = 0.28]{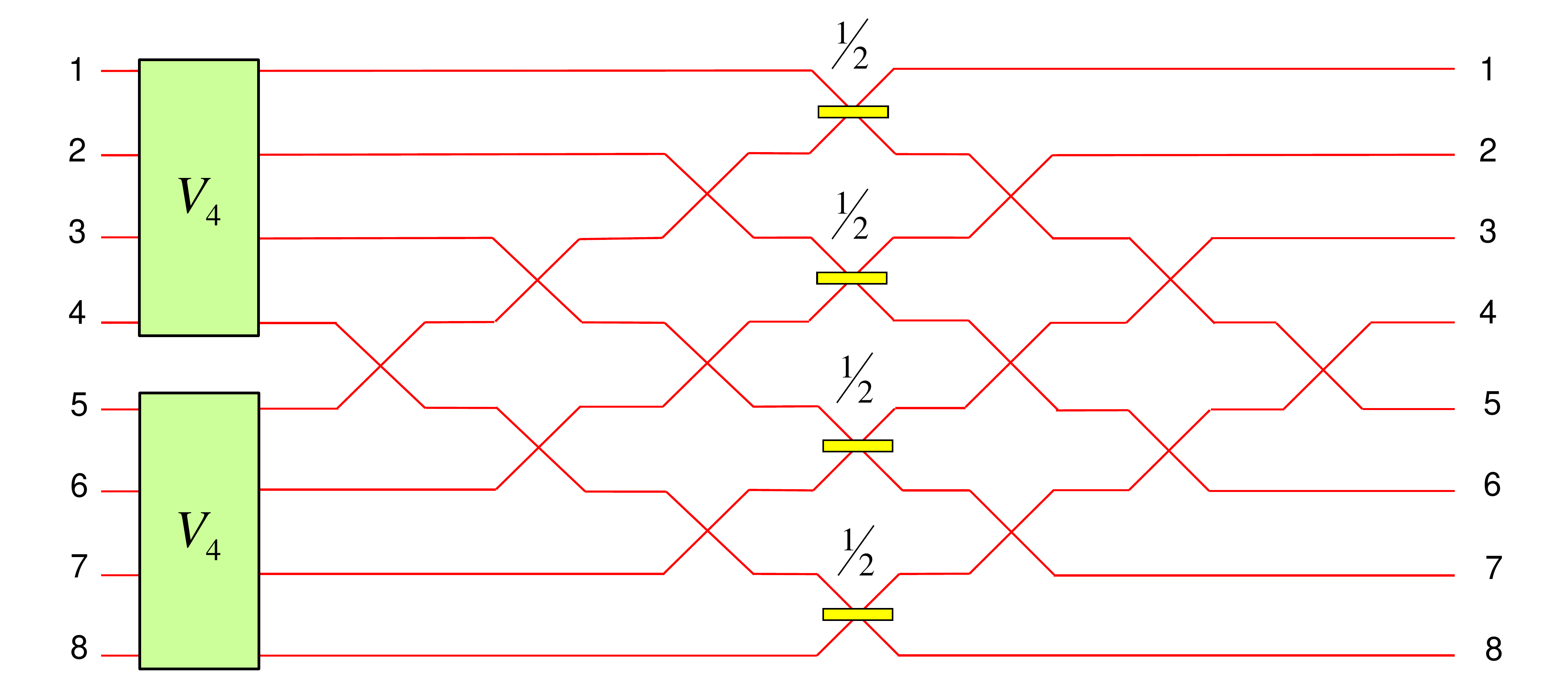}
\caption{(Color online) Implementing the unitary $V_8$
using two multiport circuits for $V_4$.}
\label{fig.groverIterV8}
\end{figure*}

From $V_4$, we can build any other circuit $V_d$ for $d = 2^k$
modes according to the rule
\begin{align}
\nonumber
V_{2d}&(1, 2,\ldots, 2d)
    = [V_d(1,\ldots, d) \> V_d(d+1,\ldots, 2d)] \\
\nonumber
&\qquad [\Sigma(1,2, \ldots, 2d)]
[B(1,2) \> B(3,4) \cdots B(2d-1,2d)] \\
&\qquad \left[\Sigma^{-1}(1,2, \ldots, 2d)\right],
\end{align}
where $\Sigma$ is the same shuffle operator defined
in Eq.~(\ref{eq.shuffleSigma}).
As an example, the circuit for $V_8$ is shown in
Fig.~\ref{fig.groverIterV8} .

\begin{figure}[t]
\centering
\includegraphics[scale = 0.3]{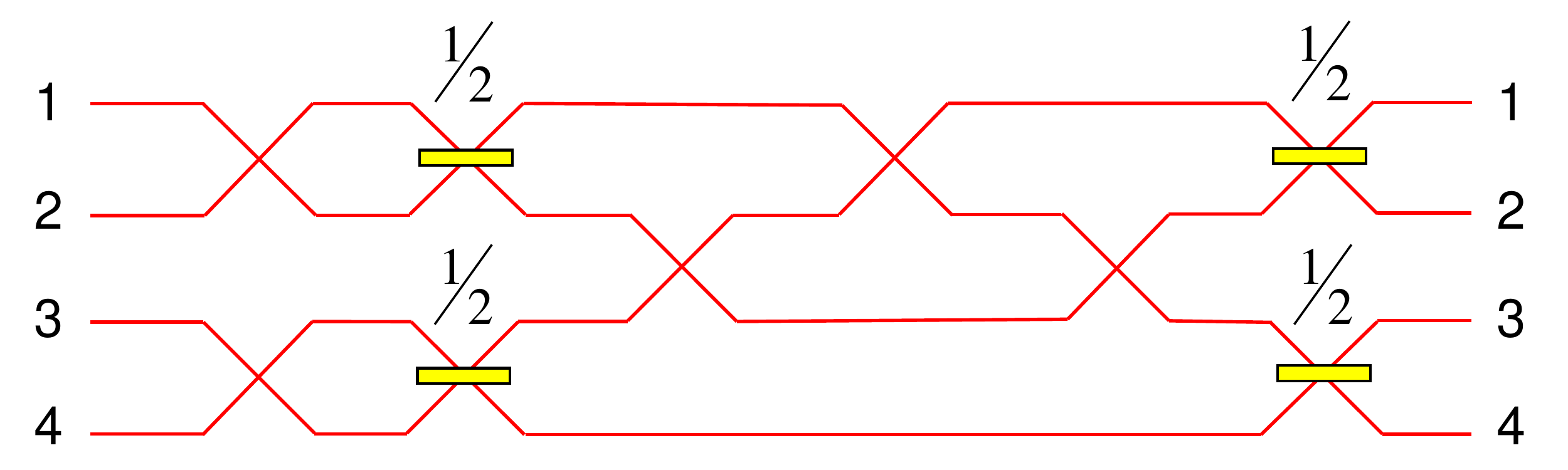}
\caption{(Color online) The optical multiport circuit for Grover
  inversion $W_4$. }
\label{fig.groverIterDim4}
\end{figure}

Let $W_d$ denote Grover inversion on $d$ modes.
First let us consider $W_4$, which is given by
\begin{align}
\nonumber
W_4&(1,2,3,4) = [S(1,2) \> S(3,4)] \> [B(1,2)\> B(3,4)] \> [S(2,3)]   \\
&\qquad  [S(1,2)] \> [S(2,3)] \> [B(1,2) \> B(3,4)]  .
\end{align}
Fig.~\ref{fig.groverIterDim4} shows how $W_4$
can be implemented with integrated optics.

In the general case, it is convenient to define
the unitary operation $\Phi$ on $2d$ modes that
exchanges the photon amplitude in mode $1$ and mode $d+1$
through the following network of swap gates on neighboring modes:
\begin{align}
\nonumber
\Phi &(1,2,\ldots,2d) = [S(d,d+1)] \\
\nonumber
&\>   [S(1,2) \> S(3,4)  \cdots S(d-1,d)] \\
\nonumber
&\>  [S(2,3) \> S(4,5)  \cdots S(d-2,d-1)]  \cdots \\
\nonumber
&\> \left[S\left(\frac{d}{2},\frac{d}{2}+1\right) \right]
 \left[S\left(\frac{d}{2}-1,\frac{d}{2}\right)
\> S\left(\frac{d}{2}+1,\frac{d}{2} + 2\right) \right]  \\
&\>   \cdots [S(1,2) \> S(3,4)  \cdots S(d-1,d)] [S(d,d+1)].
\end{align}
It is worth mentioning that $\Phi(1,2\ldots, 2d)$ employs
a total of $\frac{d^2}{4} +\frac{d}{2}+1$ swap gates.

\begin{figure*}[t]
\centering
\includegraphics[scale = 0.28]{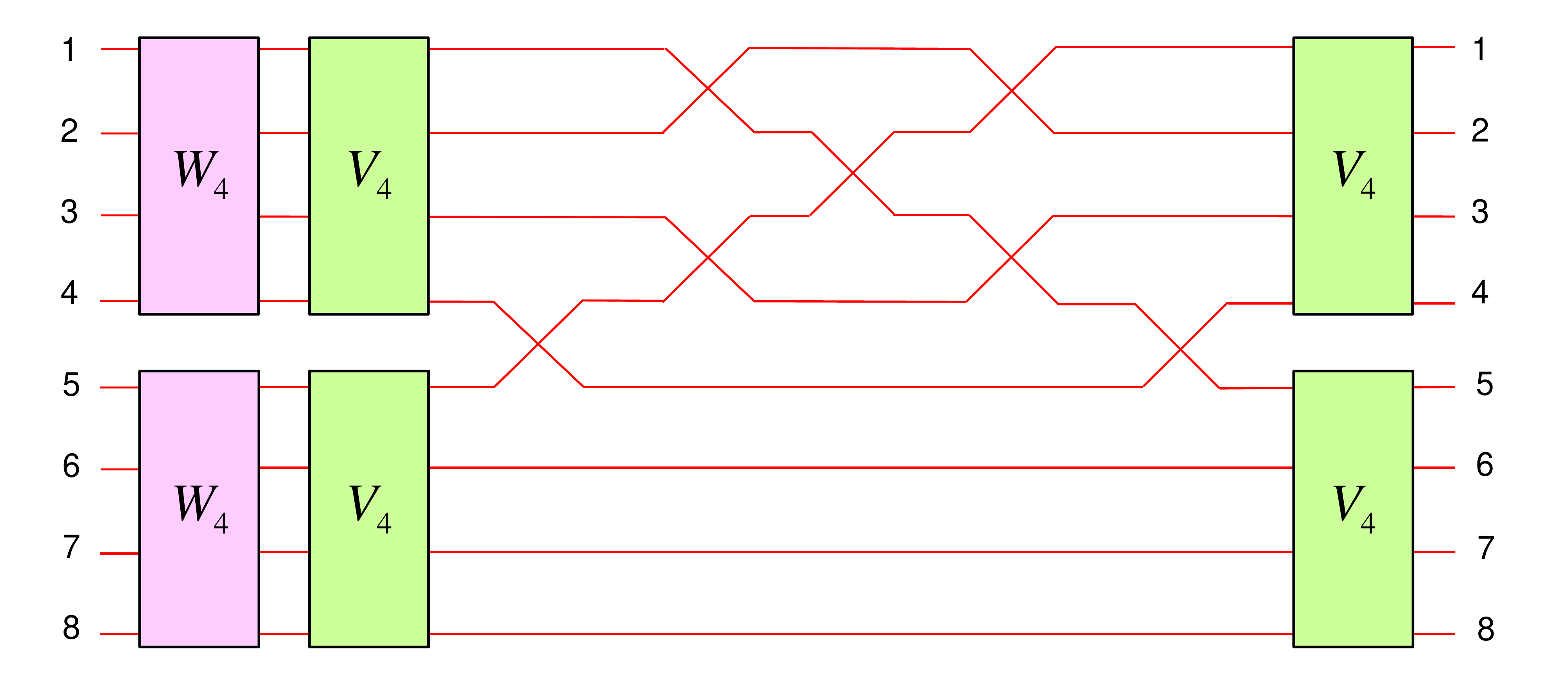}
\caption{(Color online) Implementing Grover inversion $W_8$ using pairs of
multiport circuits for $W_4$ and $V_4$.}
\label{fig.groverIterDim8}
\end{figure*}

The general multiport circuit for Grover inversion $W_{2d}$
employs both $W_d$ and $V_d$, and is given by
\begin{align}
\nonumber
W_{2d}&(1,2,\ldots, 2d) = \left[W_d(1,\ldots d)
     \> W_d(d+1,\ldots, 2d)\right] \\
\nonumber
&\> [V_d(1,\ldots d) \> V_d(d+1,\ldots, 2d)]
     [\Phi(1,2,\ldots, 2d)] \\
&\> [V_d(1,\ldots d) \> V_d(d+1,\ldots, 2d)].
\end{align}
Fig.~\ref{fig.groverIterDim8} illustrates how
the rule is used for building the circuit for $W_8$.

Counting the number of optical elements used in $W_{2d}$,
first for the unitary $V_d$, we have
\begin{equation}
N\left(V_{2d}\right) = 2N(V_d) + d(d-1) + d,
\end{equation}
where the first term refers to a pair of $V_d$ circuits,
the second term refers to the elements in $\Sigma$ and
$\Sigma^{-1}$, and the last term refers to a set of
parallel equal beam splitters on neighboring modes.
Solving the formula with $N\left(V_4\right) = 6$, we obtain
\begin{equation}
N\left(V_d\right) = \frac{d(d-1)}{2}.
\end{equation}

Now for Grover inversion, we have
\begin{equation}
N\left(W_{2d}\right)=2 N\left(W_d\right) + 4 N\left(V_d\right)
   +d^2+d+1,
\end{equation}
where $N(W_4) = 9$.
Plugging in $N\left(V_d\right)$ and solving the relation
yields
\begin{equation}
N\left(W_d\right) = \frac{9d^2 -d\left(6\log_2{d} + 4 \right)}{8} -1.
\end{equation}

The recursive scheme directly implies that the $W_d$ has a
matrix decomposition given by
\begin{align}
W_{2d} &=
\begin{pmatrix}
	V_d & 0 \\
	0 & V_d
\end{pmatrix}
Q
\begin{pmatrix}
	V_d & 0 \\
	0 & V_d
\end{pmatrix}
\begin{pmatrix}
	W_d & 0 \\
	0 & W_d
\end{pmatrix}, \\
V_{2d} &= \left(H \otimes I_d \right)
\begin{pmatrix}
	V_d & 0 \\
	0 & V_d
\end{pmatrix},
\label{eq:grovDecomp}
\end{align}
where $I_d$ is the $d$-dimensional identity matrix,
\begin{align}
W_2 &=
\begin{pmatrix}
	0 & 1 \\
	1 & 0
\end{pmatrix},
&
V_2 &= H = \frac{1}{\sqrt{2}}
\begin{pmatrix}
	1 & 1 \\
	1 & -1
\end{pmatrix},
 \end{align}
and $Q$ is the $2d \times 2d$ permutation matrix that
exchanges the first and $(d+1)$th entry of a column vector.
Unlike in the QFT circuit, the construction for Grover inversion
is known only to work when $d = 2^k$, because we do not know of
any natural counterpart to $V_d$ when $d$ is not a power
of two.

\section{Multiport circuit simulations}
\label{sec:mCircSim}

In realistic linear optical systems, optical elements experience
photon losses, optical modes suffer from relative phase mismatches,
and fabrication defects lead to errors in the splitting ratios
of beam splitters. To account for such device imperfections,
we follow the example of Ref. \cite{mower2014} and consider a
simple model for silicon-based PICs that use
directional couplers \cite{mikkelsen2014} for beam splitters and
thermo-optic phase modulators \cite{harris2014} for phase shifters.

In this section, we simulate experiments on our multiport circuits
and assess their performance under a fabrication model that
focuses on two primary sources of errors: (i) incorrect reflectivities
in beam splitters and (ii) absorption losses in the phase shifters.

For directional couplers, the error in the splitting ratio is due
to imperfections in the dimensions of the coupled waveguides.
In our model, we assign reflectivities to the beam splitters in
the multiport circuits according to a Gaussian distribution,
with mean 0.5 and standard deviation 0.04, which agrees with the
testing data on recently developed devices \cite{mikkelsen2014}.

For thermo-optic modulators, the important source of error is the
free-carrier absorption in doped silicon material, leading to
propagation loss. Typically, the absorption process is modeled
as a unitary operation by introducing a beam splitter between
the lossy mode and an ancillary one, and whose transmissivity
represents the photon loss rate for the device.
Fortunately, the matrix representing a phase
shifter is always block diagonal in the circuit. It is therefore
sufficient to apply a scaling factor $\sqrt{1-\gamma}$ on the
lossy mode, where $\gamma$ is the absorptivity.
In our model, we assign $\gamma$ values to the phase shifters
according to a rectified Gaussian distribution, with mean $0.05$
and standard deviation $0.025$, consistent with the reported loss
rates in the latest studies \cite{harris2014}.

Our scheme also utilizes a significant number of swap gates,
each of which can be realized by a beam splitter with vanishing
reflectivity. However, it may be possible to implement
the swap more efficiently since there is no required interaction
between the modes. As such, we chose to model them as slightly
better performing beam splitters, whose reflectivities are drawn
from a rectified Gaussian distribution of mean 0.02 and standard
deviation 0.02.

The first experiment we considered involves implementing the QFT
circuit with randomly generated inputs of the form
\begin{align}
\nonumber
\ket{\phi_4} &= \left( z_1, z_2, z_3, z_4 \right)^T / \zeta_4
\end{align}
in the $2$-qubit case and
\begin{align}
\nonumber
\ket{\phi_8} &= \left( z_1, z_2, z_3, z_4, z_5, z_6, z_7, z_8 \right)^T
/ \zeta_8
\end{align}
in the $3$-qubit case where
$z_i = \sqrt{-2 \ln x_i} \exp\left(2\pi y_i\right)$
and $\zeta_j = \sum_{i=1}^j |z_i|^2$.
The parameters $x_i$ and $y_i$ were all picked independently and
uniformly at random from the interval $(0, 1)$.

It is known that a Haar-random $d$-dimensional pure state may be
constructed by normalizing a vector of $d$ independent, identically
distributed complex Gaussian random variables \cite{zykowski2001}.
Thus, the above procedure for generating input states is equivalent
to sampling random pure quantum states from the uniform Haar measure,
since $z_i$ represents the Box-Muller transform \cite{boxmuller1958},
which generates a pair of Gaussian random numbers from a
pair of uniformly distributed ones.

To evaluate the performance of our circuits, we calculated the
(squared) fidelity
$F(\Psi, \Phi) = |\bracket{\Psi}{\Phi}|^2$
between the simulated output state $\ket{\Phi}$ and
the ideal output $\ket{\Psi}$ for each particular input.
Because of the Haar-random sampling of input states,
the mean fidelity we get provides a good estimate of
the average gate fidelity for the QFT circuit, a fairly common
figure of merit for quantum gate experiments.

After running $10^7$ trials, we obtained fidelities with
a mean value of $0.94$ and a standard deviation of $0.032$
in the 2-qubit case, and a mean value of $0.86$ and a
standard deviation of $0.056$ in the 3-qubit case.

The second experiment involves running Grover's algorithm on
four-item and eight-item databases, each with a unique solution.
In contrast with the first experiment, Grover's algorithm
specifically uses the equal superposition state as its input.
Thus, we have included the gates required for preparing this
state in our simulation.

\begin{figure}[t]
\centering
\includegraphics[scale = 0.3]{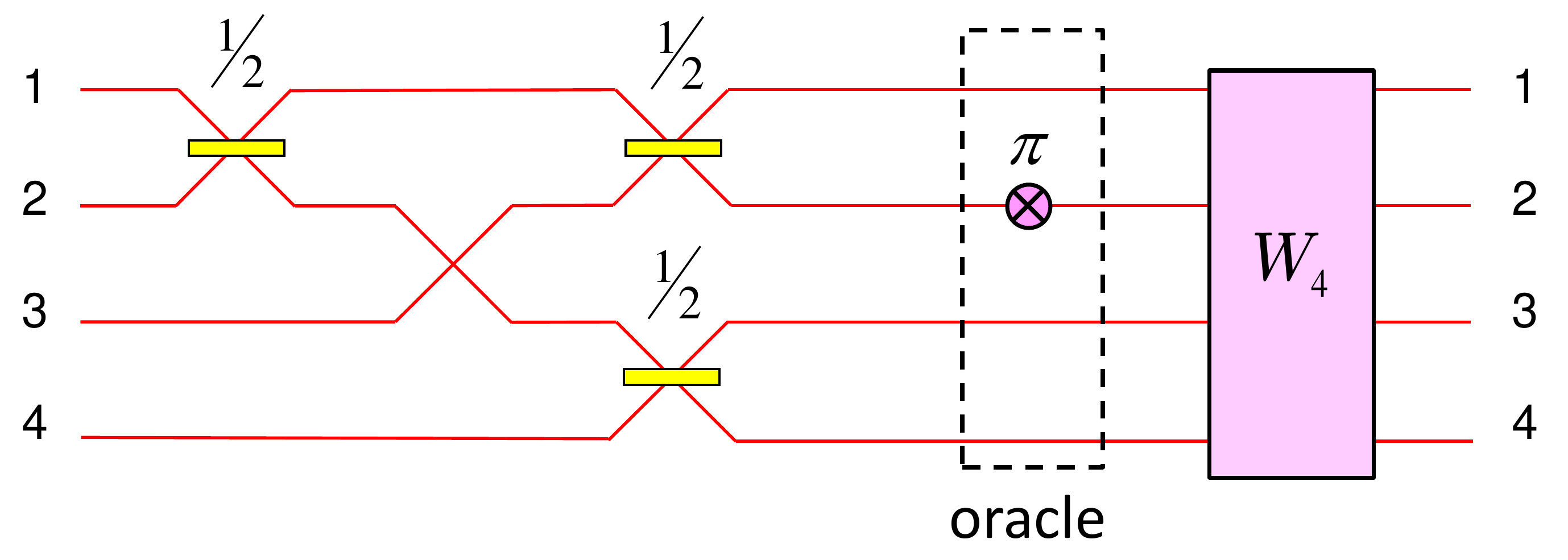}
\caption{(Color online) The optical multiport circuit for Grover search
 on four items. In the diagram, the second mode represents the
 solution to the search problem. The output of the algorithm is
 obtained by a final measurement in the standard basis.}
\label{fig.groverAlg4}
\end{figure}

\begin{figure}[t]
\centering
\includegraphics[scale = 0.28]{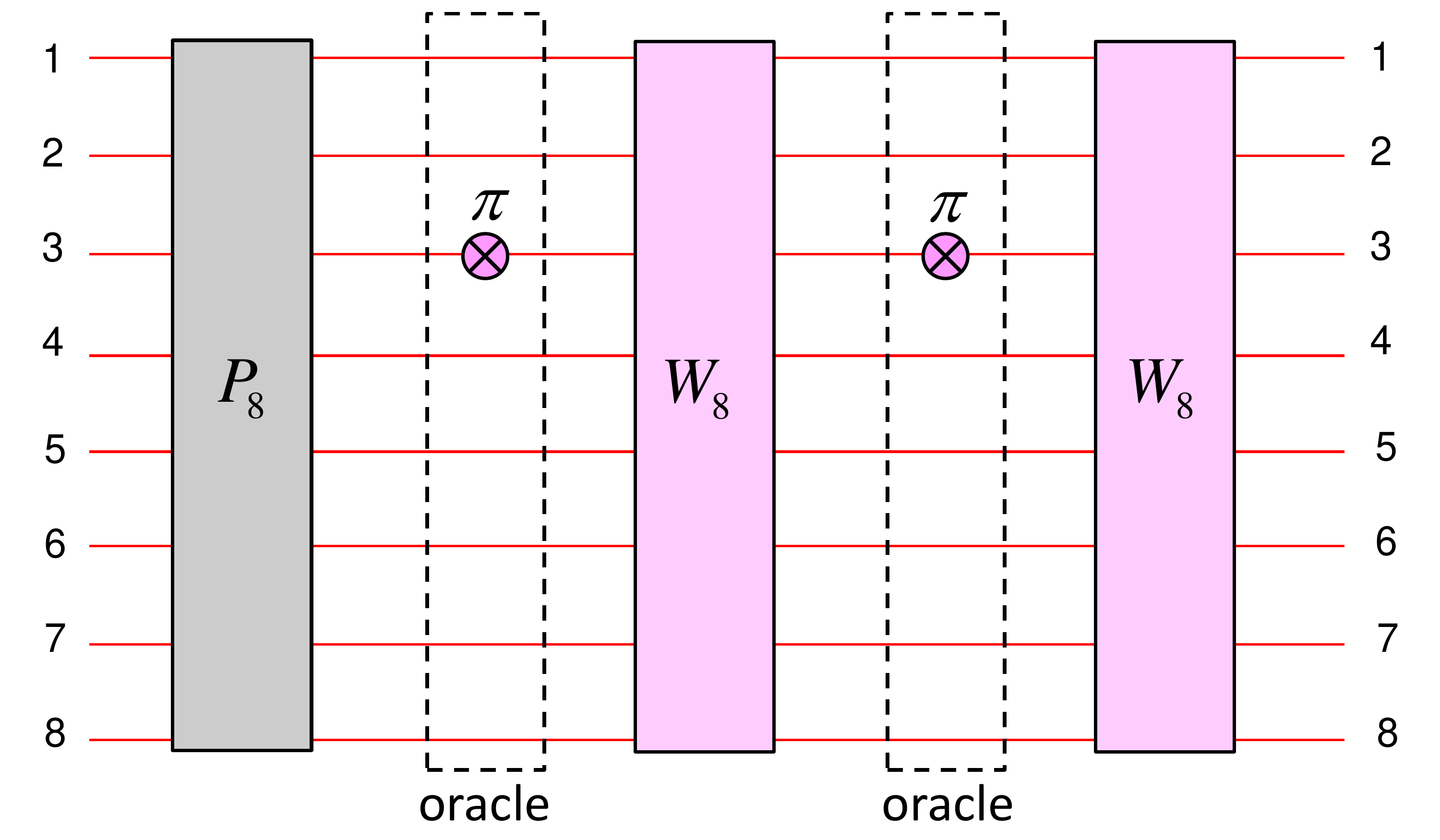}
\caption{(Color online) The optical multiport circuit for Grover search
   on eight items. The box $P_8$ corresponds to the circuit
	in Fig.~\ref{fig.groverInput8}. In this diagram, the
	third mode represents the solution to the search problem.
  The output is obtained by a final measurement in
  the standard basis.}
\label{fig.groverAlg8}
\end{figure}

\begin{figure}[t]
\centering
\includegraphics[scale = 0.28]{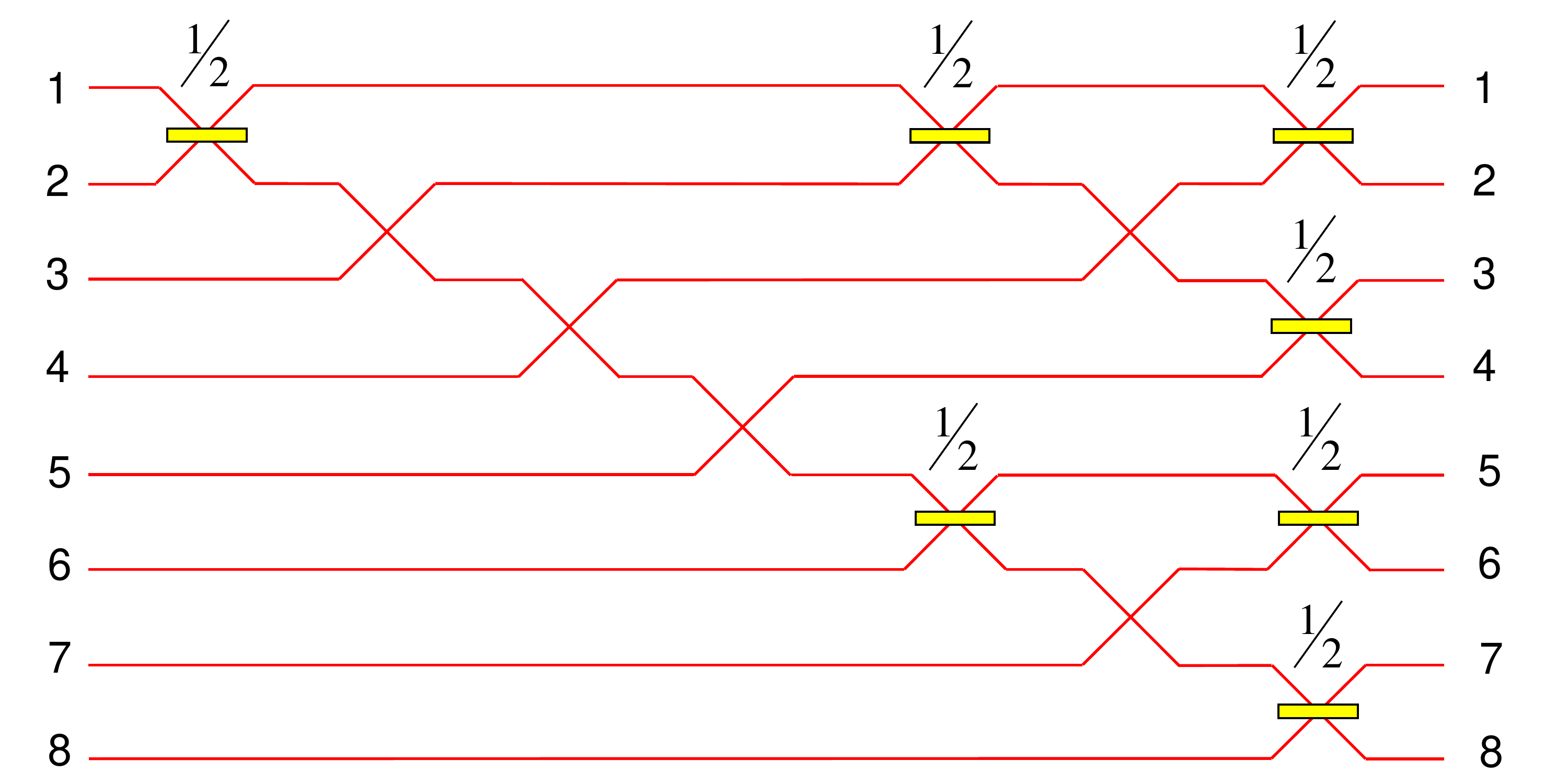}
\caption{(Color online) The optical multiport circuit $P_8$ for preparing
 an equal superposition state for eight modes
 given an input photon in the first mode.}
\label{fig.groverInput8}
\end{figure}

The multiport circuit for a four-item Grover search is illustrated
in Fig.~\ref{fig.groverAlg4}. The schematic for the eight-item
version is displayed in Fig.~\ref{fig.groverAlg8}, where
the dashed boxes refer to oracle queries realized by a $\pi$-phase
shift on the appropriate mode, and the circuit $P_8$ for preparing
the input is shown separately in Fig.~\ref{fig.groverInput8}.

Observe that the oracle query and Grover inversion are repeated
twice in the $8$-item quantum search, which follows from the fact
that Grover's algorithm prescribes doing
$\left\lfloor \frac{\pi}{4} \sqrt{d} \right\rfloor $
iterations, which is 2 when $d = 8$.
In this case, Grover's algorithm produces an output state that
generates the solution with high probability
$\left(\frac{121}{128} \approx 0.945 \right)$, not
deterministically.

We performed $10^7$ trials for both searches, each with a
randomly selected unique solution. For each trial we
computed the fidelity of the simulated output with the ideal one.
We obtained fidelities with a mean of $0.90$ and a
standard deviation of $0.051$ for a four-item Grover search, and
a mean of $0.76$ and a standard deviation of $0.099$
for an eight-item Grover search

We must emphasize that what these results indicate is the
feasibility of a classical simulation of Grover's algorithm
with our circuits.  Nevertheless, like QFT, it is
possible that the unitary $W_d$ may find
suitable application in some other truly quantum setting.

There have been no linear optical experiments to date to
which we can directly compare our simulation values; however,
recent demonstrations on PIC devices have obtained fidelities
of more than 0.94 \cite{politi2008, carolan2015}.
Considering that our simulations were done with a rudimentary
yet conservative error model, it is encouraging to see
a comparable level of performance with our multiport circuits.

\begin{table}[t]
  \begin{tabular}{ c c c }
  \hline \hline
  Experiment & Mean fidelity & Std. deviation \\
  \hline
  2-qubit QFT [8] & 0.944 & 0.0319 \\
  4-item Grover  [14] & 0.904 & 0.0507 \\
  3-qubit QFT [41] & 0.861  & 0.0559 \\
	8-item Grover  [112] & 0.762 & 0.0990 \\
  \hline     \hline
  \end{tabular}
 \caption{Simulation fidelities for $10^7$ trials. The total number of
beam splitters and phase shifters in each circuit is indicated in the brackets.}
 \label{tab.simFidelity}
\end{table}

\section{Discussion}

\begin{figure*}[t]
\begin{subfigure}{.5\linewidth}
  \centering
  \includegraphics[width=.65\linewidth]{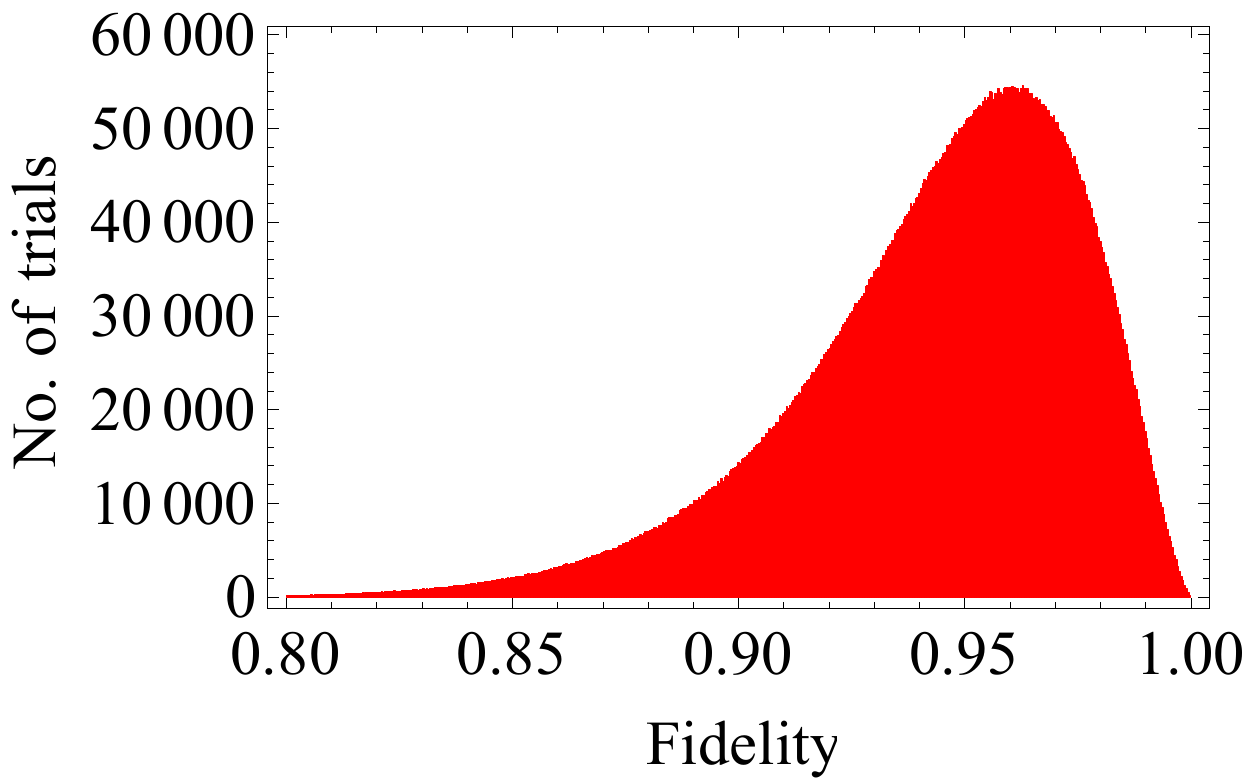}
  \caption{}
  \label{fig:sfiga}
\end{subfigure}%
\begin{subfigure}{.5\linewidth}
  \centering
  \includegraphics[width=.65\linewidth]{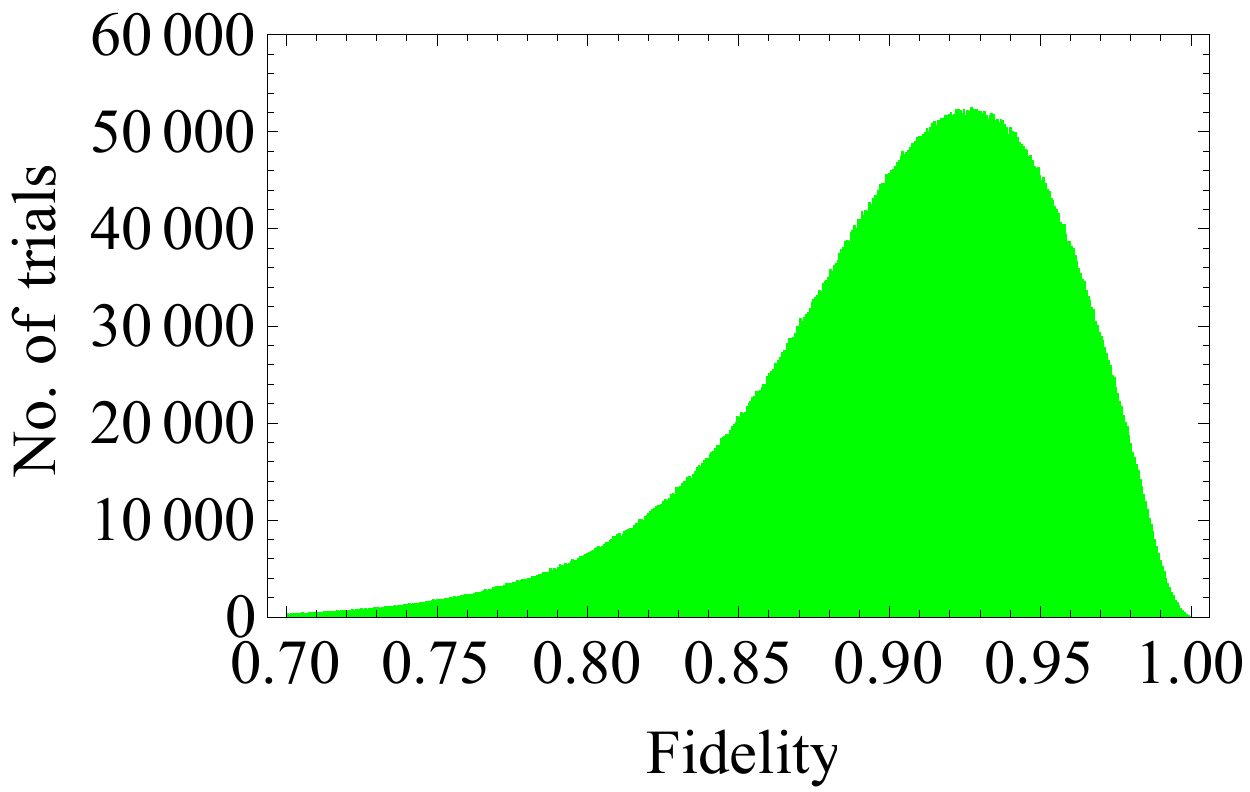}
  \caption{}
  \label{fig:sfigb}
\end{subfigure}
\begin{subfigure}{.5\linewidth}
  \centering
  \includegraphics[width=.65\linewidth]{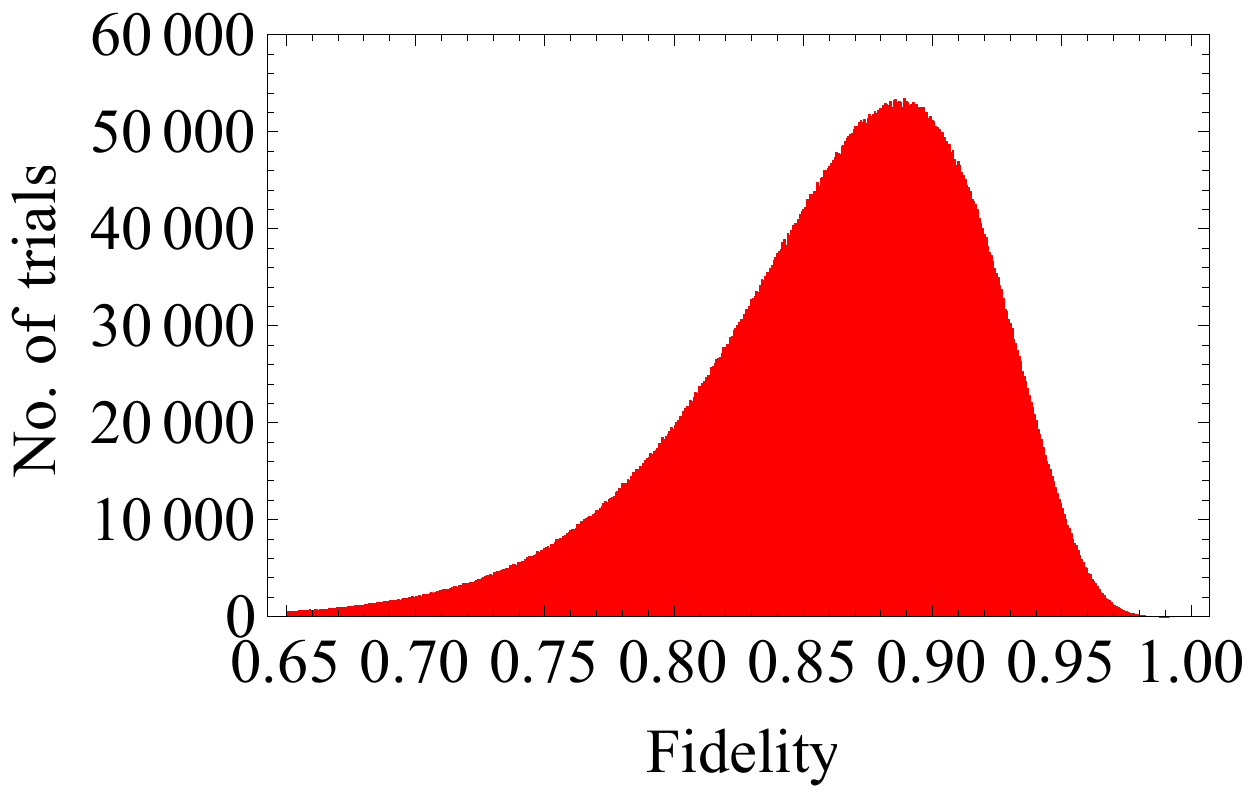}
  \caption{}
  \label{fig:sfigc}
\end{subfigure}%
\begin{subfigure}{.5\linewidth}
  \centering
  \includegraphics[width=.65\linewidth]{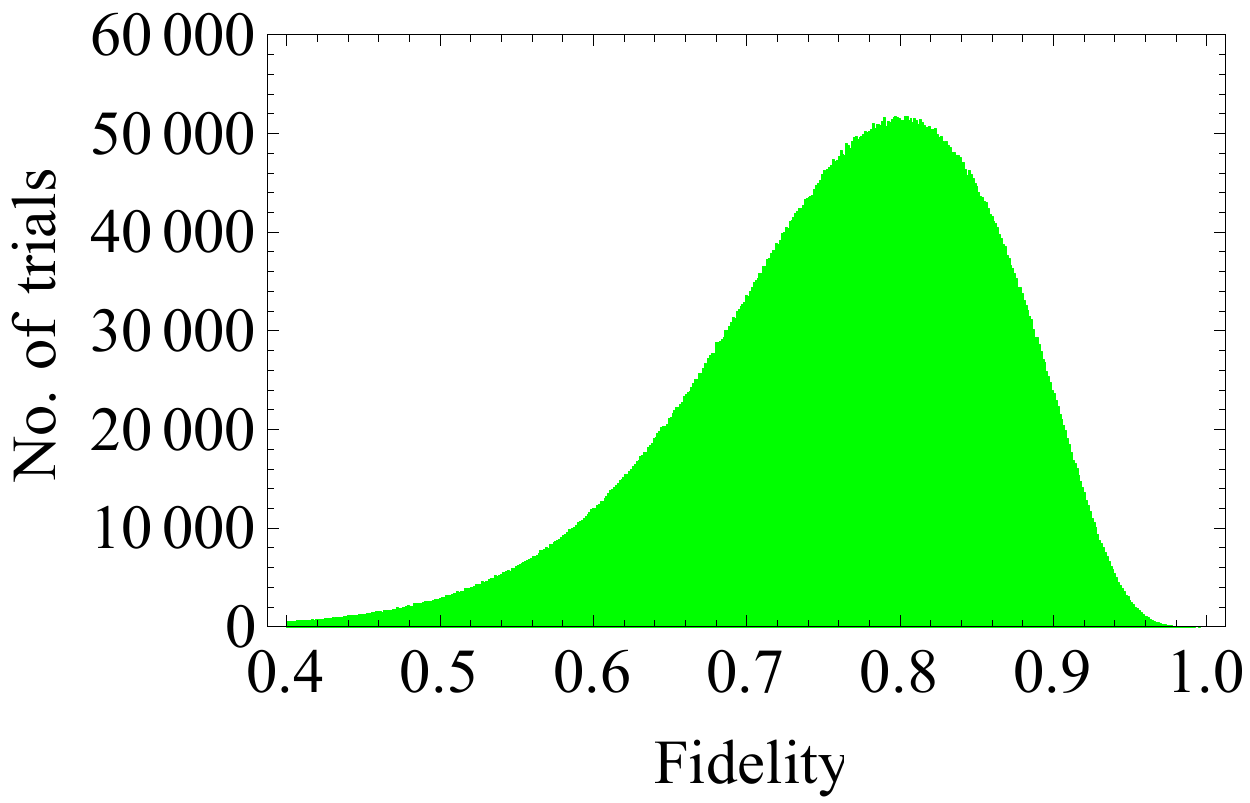}
  \caption{}
  \label{fig:sfigd}
\end{subfigure}
\caption{(Color online) Histograms of simulation results for (a) $2$-qubit QFT ($0.949$),
(b) four-item Grover search ($0.912$), (c) $3$-qubit QFT ($0.870$), and
(d) eight-item Grover search ($0.774$), with bin sizes obtained using the
Freedman-Diaconis rule . The median values are indicated in parentheses.}
\label{fig.simPlots}
\end{figure*}

Any $d$-dimensional unitary operation can be implemented with linear
optics, by a triangular array of beam splitters on $d$ modes, each
accompanied by a phase shifter, and another $d-1$ phase
shifters \cite{reck1994}. Such a matrix decomposition requires
$d^2-1$ optical elements, exactly the number of real parameters
needed to specify a unitary in $\mathrm{SU}(d)$.
While this specific arrangement might be preferred when
trying to develop a fully reconfigurable linear optical processor,
it is generally not the optimal choice for specific unitary families.
If we wish to develop a quantum processor dedicated to a particular
task, it may be possible, and quite desirable, to design a multiport
circuit that uses fewer elements, since fabrication defects in
devices are known to severely hamper the quantum performance of PICs.
Furthermore, depending on how technological capabilities improve,
photonic circuits that consist just of equal beam splitters and
a small number of simple phase shifts might be more appropriate
in certain contexts.

The recursive scheme described in this paper illustrates that for
unitary operations that serve as major components of Shor's and
Grover's algorithm, we can assemble a multiport circuit from smaller
versions of the same operation. The matrix decompositions for QFT
and Grover inversion both have a relatively simple structure,
which is made seemingly complicated only because of permutations
have to be implemented by a sequence of nearest-neighbor swap gates.

We have previously discussed the optical circuit size for QFT
and Grover inversion. For completeness, we can also remark on the
optical circuit depth in comparison to the unitary matrix
factorization of  Reck, \emph{et al.} Let $D(U)$ denote the circuit depth
for implementing $U$. In their case, the configuration of beam
splitters and phase shifter is fixed for any arbitrary
$d$-dimensional unitary so the circuit depth is $2(d-1)$ .
For the QFT circuits, we have
\begin{equation}
D(F_d) = 3(d-1) - 2 \log_2 d
\end{equation}
while for the Grover inversion circuits, we have
\begin{equation}
D(W_d) = 5d - 2 \log_2 d - (\log_2 d)^2 - 6.
\end{equation}
Thus, there is a linear dependence on $d$ in all cases, although
our circuit depths are around a factor of 2 worse. This is
precisely due to the swap operations in our scheme, which is
quite unavoidable because elementary gates on a PIC are restricted
to adjacent optical modes.

The simulation results listed in Table~\ref{tab.simFidelity}
show that these circuits perform well even when realistic
errors in the optical elements are taken into account.
This is accomplished even before we apply post-fabrication
optimization techniques \cite{mower2014} that would significantly
improve the overall performance.

For a quick summary of the data, we include histograms of the
fidelities in Fig.~\ref{fig.simPlots}, where the bin size is
determined by the Freedman-Diaconis rule. This gives a rough
picture of how the fidelity values are distributed, in particular,
how much are the values skewed to the left.

Nonetheless, it seems clear that further advances in fabrication
methods will be needed to scale to much larger circuits.
From the results of both experiments, we can roughly estimate a
9\% to 15\% reduction in average fidelity when the number of modes
is doubled. This seems to be the case even though the number of
elements in QFT increases fivefold while that of Grover's algorithm
increases eightfold. Still, this rate of decline implies
the fidelity will go below acceptable levels rather quickly.
It is previously known that errors of as high as 1\% per elementary
gate will still allow fault-tolerant quantum computing \cite{knill2005}
but in terms of the overhead resources needed to protect against
failures, we will likely require less than 0.1 \% error per gate
for a practical device.

There are other long-standing issues with integrated photonic
platforms. One is the problem of integrating single photon sources
and detectors with a silicon-based PIC. Some recent progress in
this area include waveguide-integrated semiconductor quantum
dot sources \cite{murray2015} and superconducting
nanowire detectors \cite{najafi2015}.

Another problem is that one typically needs to build customized
chips for each experiment. Furthermore, settings may need to be
reconfigured between different runs of the same experiment;
for example, in an adaptive experiment, several device settings might
be updated according to the outcome of intermediate measurements.
These latter concerns will be addressed by a reprogrammable
photonic quantum processor \cite{mower2014, carolan2015}.

\section{Concluding Remarks}

Photonic approaches to quantum information processing has seen
extensive growth over the last few years and linear optics has
played a major role in many of the recent advances. In particular,
optical implementations on integrated systems have vastly improved
our ability to perform quantum experiments with single photons.

In this paper we described recursive multiport circuits for quantum
Fourier transforms and Grover inversion. The circuits are designed
to be implemented on linear optical networks where qudits are encoded
in the path traversed by single photons. Each circuit requires
$O(d^2)$ optical devices for implementing a $d$-dimensional unitary
operation, which shows that they scale in a reasonable manner.
We also demonstrated that these circuits can achieve high-fidelity
performance in a practical setting by conducting simulations on
a silicon-based PIC model that incorporates the faulty operation
of real optical devices.

Our recursive schemes take advantage of the natural
modularity of photonic integrated circuits, allowing us
to construct more sophisticated multiport circuits using
smaller versions of the same operation as building blocks.
This may prove beneficial for practical implementations
because it provides a straightforward approach to scaling operations
into higher dimensions. In contrast, if one for example uses
the triangular array of Reck, \emph{et al.}\cite{reck1994}, the phases
and reflectivities in the circuit will generally require adjusting
whenever more optical modes are added.

Our results show the versatility of PICs as a platform
for developing high-quality, scalable quantum information processors.
It is our hope that this work will help promote new avenues of
research into integrated photonic devices and stimulate technological
improvements that will allow more practical applications.

\begin{acknowledgments}
This work is funded by institutional research grant IUT2-1
from the Estonian Research Council and by the European Union through
the European Regional Development Fund.
\end{acknowledgments}


\bibliography{recursiveQCirc}

\end{document}